\documentclass[10pt]{article}
\usepackage{latexsym}
\usepackage{amsmath}

\long\def\ca#1\cb{} 

\newcommand{\becs}{\begin{cases}}
\newcommand{\bem}{\begin{matrix}}

\newcommand{\encs}{\end{cases}}
\newcommand{\enm}{\end{matrix}}

\def\outl#1{\par{\medskip\noindent\hspace*{.5cm}\bf
      \mathversion{bold}#1\mathversion{normal}\smallskip} }
 \def\xa{} \def\xb{}  

 \def\outl#1{}  \def\xa{} \def\xb{}  

\ca
 \def\outl#1{\par{\medskip\noindent\hspace*{.5cm}\bf
      \mathversion{bold}#1\mathversion{normal}\smallskip} }
 \long\def\xa#1\xb{}
 
\cb

\begin{document}

\title{Reply to Comment on ``Particle Path Through a Nested Mach-Zehnder
  Interferometer'' }
\author{Robert B. Griffiths\thanks{Electronic address: rgrif@cmu.edu}\\
Department of Physics,
Carnegie Mellon University,
Pittsburgh, PA 15213}
\date{Version of 9 February 2018}
\vspace*{-2.4cm}

{\let\newpage\relax\maketitle}

\vspace{-1cm}

\begin{abstract} 	
  While much of the technical analysis in the preceding Comment \cite{Slh18} is
  correct, in the end it confirms the conclusion reached in my previous work
  \cite{Grff16}: a consistent histories analysis provides no support for the
  claim of counterfactual quantum communication put forward in \cite{Slao13}
\end{abstract}

In \cite{Slao13} Salih, Li, Al-Amri and Zubairy presented a quantum optics
protocol, hereafter referred to as SLAZ, which, they claimed, makes possible
\emph{counterfactual communication}: information can be transmitted from Bob to
Alice without any photon being present in the optical communication channel
that connects them. Vaidman in \cite{Vdmn14} (and also in some subsequent work
\cite{Vdmn15,Vdmn16}) disputed this claim, to which Salih et al.\ responded in
\cite{Slao14}. Later the counterfactual claim, in particular as presented in
\cite{Slao14}, was criticized in Sec.~VII of \cite{Grff16} from the perspective
of the \emph{consistent histories} (CH) interpretation of quantum mechanics.
The Comment \cite{Slh18} by Salih that precedes this Reply challenges the
correctness of the analysis in \cite{Grff16} by presenting an alternative CH
analysis of the SLAZ protocol. While the analysis in \cite{Slh18} of certain
consistent families is correct, I will argue that this does not undermine the
criticism in \cite{Grff16}. That is to say, the possibility of counterfactual
quantum communication has not been demonstrated, and instead the CH analysis in
\cite{Slh18} confirms the presence of a serious problem with the SLAZ protocol.

For readers unfamiliar with the CH approach we note that it is an
interpretation of standard quantum mechanics---Hilbert space, Schr\"odinger
equation, Born rule, no hidden variables---using sequences of events at
successive times (``histories'') represented by a series of projectors on
subspaces of the quantum Hilbert space. A collection of mutually exclusive
histories, one and only one of which will occur during a particular run,
constitutes a \emph{family}, and probabilities can be assigned to individual
histories in such a family using an extended version of the Born rule provided
certain \emph{consistency conditions} are satisfied, making it a
\emph{consistent family of histories}. A short introduction to CH will be found
in \cite{Grff14b}, a detailed application to the type of problem considered
here in the earlier sections of \cite{Grff16}, while \cite{Grff02c} is a
standard reference. Unlike textbook quantum theory, ``measurement'' does
\emph{not} play a fundamental role in CH; physical measurements as carried out
in the laboratory are simply instances of quantum processes governed by
fundamental physical principles that apply to all quantum systems with no
special role for measurements.

The technical results in \cite{Slh18} obtained using the CH approach are
correct. For the first outer cycle of the SLAZ protocol, Fig.~2 of
\cite{Slh18}, there is a consistent family which allows for the possibility
that the photon is in the communication channel $C$ connecting Alice and Bob at
times $t_2$ and $t_3$, and shows that the probability of finding it there is 0
(assuming it is not absorbed at the end of the cycle), consistent with the
claim of counterfactual communication. The same is true of the second cycle
when considered in isolation. However, when the output of the first cycle is
fed into the second cycle and the two are considered together in sequence, the
consistency conditions for a family that includes the possibility that the
photon is or is not in the $C$ channel at each of the four successive times
when it could be there, are \emph{not} satisfied, so this family is not
consistent. Incidentally, this calculation shows that the two outer cycles
in series \emph{cannot} be thought of as two successive runs of an experiment
in the sense intended (but perhaps carelessly stated) in Sec.~16.4 of
\cite{Grff02c}, as quoted in \cite{Slh18}.

Thus in his Comment Salih has demonstrated the inconsistency of the SLAZ
protocol if there are two outer cycles, and adding additional outer cycles
will not improve the situation: there will be no consistent family in which it
makes sense to say that at each of the relevant times the photon was not in the
communication channel $C$. Hence the statement in the Abstract of \cite{Slh18}
that for the Mach-Zehnder version of the protocol, ``no family of consistent
histories exists where any history has the photon traveling through the
communication channel, thus rendering the question of whether the photon was in
the communication channel meaningless from a CH viewpoint'', applies equally to
the Michelson version with two or more outer cycles. An analysis of the
protocol from the CH perspective must take account of the collection of
successive cycles as a whole, and not just consider each outer cycle in
isolation.

In addition, one must be careful when interpreting the statement ``no family of
consistent histories exists\dots.'' When no consistent family can be
constructed that contains a particular history, both asserting that the history
occurred, or that it did \emph{not} occur, are equally meaningless from the CH
perspective. What proponents of SLAZ as a counterfactual protocol must do is
demonstrate that the probability is zero that the photon is in channel $C$ at
any time. But if that probability cannot be defined in a meaningful way, such a
demonstration is impossible, and the claim of counterfactual communication
fails. Consider as an analogy the case of two-slit interference. A history in
which the quantum particle passes through slit 2 but not slit 1 on its way to a
point of maximum constructive interference does not belong to any consistent
family, and the question of whether it passed through slit 1 is meaningless
from the CH perspective (and perhaps Feynman \cite{FyLS651} would have agreed).
But that cannot be interpreted to mean that the particle did \emph{not} pass
through slit 1. Meaningless statements cannot be true or false. To use another
analogy, if the $z$ component $S_z$ of angular momentum of a spin-half particle
is $-\hbar/2$, it is meaningless to simultaneously assign a value of
$+\hbar/2$, or $-\hbar/2$, to the $x$ component $S_x$.

Consequently, the CH analysis in \cite{Slh18} lends no support to the claim of
counterfactual communication, that the photon was never in $C$, in the SLAZ
protocol. However, a qualification is in order, since in the original paper
\cite{Slao13} the actual claim was that the probability of a photon being in
the communication channel would be negligible in the limit of a large number of
inner and outer cycles. Which is not quite the same as saying that the
probability is zero for some finite number of cycles, as in the preceding
discussion. It might be possible to find some way of estimating an average
number of times the a photon is in the communication channel during a protocol
which involves a large number of cycles, and compare it with the probability
that a bit is successfully transmitted from Bob to Alice. It is not clear (to
me) how to formulate an analysis of that sort in CH terms, which might require
some notion of ``approximate consistency'', which, while occasionally mentioned
in the CH literature, has not yet (so far as I know) been developed into a
useful tool for discussing issues of the sort considered here. In any case, the
analysis in \cite{Slh18} provides no support for the claim of counterfactual
communication.

\ca
\section*{Acknowledgments}
\cb

\end{document}